# The Critical Role of Charge Balance on the Memory Characteristics of Ferroelectric Field-Effect Transistors

Mengwei Si and Peide D. Ye, *Fellow, IEEE*

***Abstract*—Ferroelectric field-effect transistors (Fe-FETs) with ferroelectric hafnium oxide (FE HfO₂) as gate insulator are being extensively explored as a promising device candidate for three-dimensional (3D) NAND memory application. FE HfO₂ exhibits long retention over 10 years, high endurance over 10¹² cycles, high speed with sub-ns polarization switching, and high remnant polarization of 10-30 μC/cm². However, the performance of Fe-FETs is known to be much worse than FE HfO₂ capacitors, which is not completely understood. In this work, we developed a comprehensive Fe-FET model based on a charge balance framework. The role of charge balance and the impact of leakage-assist-switching mechanism on the memory characteristics of Fe-FETs with M/FE/DE/S (Metal/Ferroelectric/Dielectric/Semiconductor) gate stack is studied. It is found that the FE/DE interface and DE layer instead of FE layer is critical to determine the memory characteristics of Fe-FETs, and experimental Fe-FETs can be well explained by this model, where the discrepancy between FE capacitors and Fe-FETs are successfully understood.***

## I. INTRODUCTION

FERROELECTRIC field-effect transistors (Fe-FETs) with ferroelectric hafnium oxide [1]–[7] (FE HfO₂) as gate insulator are being extensively explored as a promising device candidate for non-volatile memory applications, especially for three-dimensional (3D) NAND memory application [8]–[11] because of the one-transistor configuration. The emerging of such tremendous interest on FE HfO₂ based Fe-FETs is because of the discovery of complementary metal-oxide-semiconductor (CMOS) compatible ferroelectricity in Si-doped HfO₂, hafnium zirconium oxide (HZO), etc. First, the fabrication process of FE HfO₂ is fully CMOS compatible. Its atomic layer deposition (ALD) based process also enables the capability for 3D integration such as in the 3D vertical NAND memory. Second,

This work is supported by SRC/DARPA JUMP ASCENT Center and Air Force Office of Scientific Research.

Mengwei Si is currently with Department of Electronic Engineering, Shanghai Jiao Tong University, Shanghai, China and was with the School of Electrical and Computer Engineering and the Birck Nanotechnology Center, Purdue University, West Lafayette, IN 47907, USA.

Peide D. Ye is with the School of Electrical and Computer Engineering and the Birck Nanotechnology Center, Purdue University, West Lafayette, IN 47907 USA (e-mail: yep@purdue.edu).

the thickness of FE HfO₂ can be scaled down to sub-5 nm [12]–[14], so that operation voltage of FE HfO₂ devices can be as low as few volts, which is much smaller than flash memory devices. The lateral dimension of FE HfO₂ devices can also be scaled to advanced technology nodes due to the ultrathin thickness. [5], [6] Third and more importantly, FE HfO₂ has a large coercive field (E$_C$) of ~ 1 MV/cm, which is about 1-2 orders of magnitude larger than conventional FE materials, such as strontium bismuth tantalite (SBT) or lead zirconium titanate (PZT), leading to a stronger immunity to depolarization effect, so that more than 10-year retention time is achieved on FE HfO₂ based Fe-FETs [3], [15], [16]. Such long retention characteristics has overcome the major obstacle of Fe-FETs based on conventional FE materials. In addition, the polarization switching time of FE HfO₂ can be as low as sub-1 ns [3], [17]–[22], which is limited by the lateral domain wall propagation, so that Fe-FETs operate faster in smaller devices, which also contribute the fast operation of ultra-scaled and high-density FE HfO₂ Fe-FETs based FE memories.

High-performance FE HfO₂ based metal/FE/metal (M/FE/M) capacitors have been demonstrated with long retention over 10 years [3], high endurance over 10¹² cycles [23], high speed with sub-ns polarization switching [3], [17]–[22], and remnant polarization (P$_R$) of 10-30 μC/cm² [14]. It is naturally expected that Fe-FET with FE HfO₂ as gate insulator would also have long retention time, high endurance and large memory window (MW) due to the long retention, high endurance and high E$_C$ and P$_R$ from FE HfO₂. However, FE HfO₂ based Fe-FETs usually cannot achieve the high memory performance predicted by the material properties of FE HfO₂. In fact, endurance on the level of only 10⁶ [5], [7], [17] or below is commonly observed on Fe-FETs with FE HfO₂ gate insulator. The memory windows of Fe-FETs with FE HfO₂ gate insulator from literature reports also have very large variations [24], which is not fully understood. Therefore, the electron transport mechanism in Fe-FETs and FE capacitors are fundamentally different. How to understand the discrepancy between FE capacitors and Fe-FETs and how to improve the MW and endurance of FE HfO₂ based Fe-FETs are crucial for FE HfO₂ based FE memory applications.

It is well known that FE HfO₂ has a high P$_R$ of about 10-30 μC/cm² [14], which is far above the maximum charge density that can be supported by conventional insulators or semiconductors. Such high P$_R$ leads to the accumulation of interfacial charge at ferroelectric/dielectric (FE/DE) interface in the gate stack of a Fe-FET [25]–[28], thereby charge balance can be achieved. All memory characteristics of Fe-FETs, such as MW, retention, endurance, speed, etc., are related with this fundamental polarization switching mechanism. [27] However, in the understanding and modeling of Fe-FETs, the charge accumulation at FE/DE interface is commonly ignored. In this work, we developed a comprehensive Fe-FET model based on this charge balance framework. The impact of charge balance and the leakage-assist-switching mechanism on the memory characteristics of Fe-FETs with M/FE/DE/S gate stack is studied. It is found that the FE/DE interface and DE layer are critical to determine the memory characteristics of Fe-FETs, and experimental Fe-FETs can be well explained by this model. The new insights in this model can successfully understand the discrepancy between FE capacitors and Fe-FETs.

## II. THE DIFFERENCE BETWEEN FE CAPACITORS AND FE-FETS

The different electron transport mechanisms in Fe-FETs and FE capacitors are because of the different structures. FE capacitors discussed here have a M/FE/M structure as shown in Fig. 1. Fe-FETs have a transistor structure, where a dielectric (DE) layer, such as SiO$_2$, is usually necessary to improve the interface quality, as shown in Fig. 2. The gate stack here is metal/FE/DE/semiconductor (M/FE/DE/S) structure. In a M/FE/M structure without external electric field (E field), the charge and potential distribution are determined by the ferroelectric polarization and the charge distribution on metal electrode due to the existence of screening length, as shown in Fig. 1. The depolarization field (E$_{dep}$) is the electric field across the FE layer. [29] In ideal case, E$_{dep}$ is small due to the highly conductive metal electrode, which is negligible if FE layer is thick. In a M/FE/DE/S structure without external electric field, the charge and potential distributions are determined by the ferroelectric polarization and voltage drop across DE layer and band bending of semiconductor. Voltage drop on metal electrode here is negligible compared to on DE layer and semiconductor. As a result, E$_{dep}$ in a Fe-FET is much larger than E$_{dep}$ in a FE capacitor. Therefore, FE capacitor has much better retention characteristics than Fe-FETs. FE HfO$_2$ has a large E$_C$ so that E$_{dep}$/E$_C$ is relatively small [16] even considering the M/FE/DE/S structure, so that long retention characteristics over 10 years can still be achieved on FE HfO$_2$ based Fe-FETs.

Remnant polarization charge density in FE HfO$_2$ is about 10-30 μC/cm$^2$. For M/FE/M structure, such high charge density can be easily compensated by metal electrodes. However, on the M/FE/DE/S, charge balance conditions cannot be satisfied in ideal case. [24], [27] First, DE layer cannot support such high charge density without breakdown. For example, the charge density in Al$_2$O$_3$ at breakdown electric field is about 7 μC/cm$^2$. Second, the maximum carrier density in common semiconductors is on the order of 10$^{13}$ /cm$^2$, corresponding to a charge density of 1.6 μC/cm$^2$, which is far below the charge balance condition. In ideal case without considering FE/DE interfacial charge, and because of the continuity of displacement field (D field), the displacement field in FE layer (D$_{FE}$) must be equal to the displacement field in DE layer (D$_{DE}$),

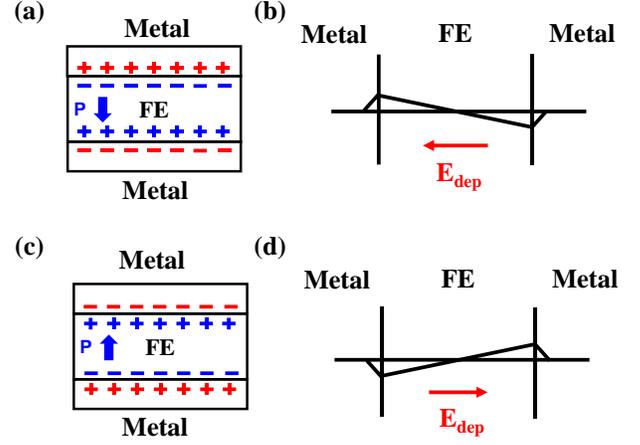

Fig. 1. (a) Charge and (b) potential distribution in a M/FE/M capacitor in polarization down state with no external electric field. (c) Charge and (d) potential distribution in a M/FE/M capacitor in polarization up state with no external electric field. The depolarization field is induced by the screening length in metal electrodes, which is usually negligible if FE layer is thick.

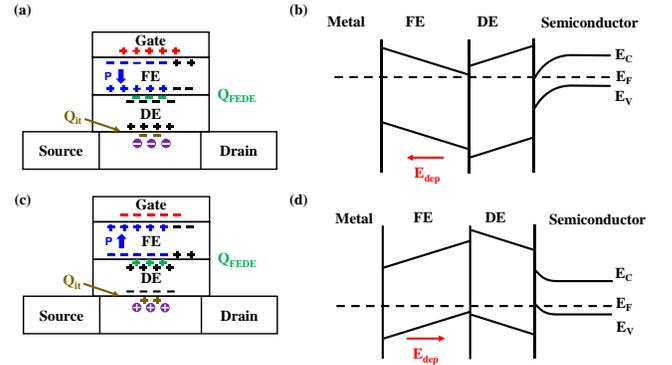

Fig. 2. (a) Charge distribution and (b) band diagram of M/FE/DE/S gate stack in polarization down state with no external electric field. (c) Charge distribution and (d) band diagram of M/FE/DE/S gate stack in polarization up state with no external electric field. The depolarization field is induced by the voltage drop across DE layer and band bending of semiconductor.

where,

$$D_{FE} = P_{FE} + \epsilon_{FE}E_{FE}, \quad (1)$$

$$D_{DE} = \epsilon_{DE}E_{DE}, \quad (2)$$

so that

$$P_{FE} + \epsilon_{FE}E_{FE} = \epsilon_{DE}E_{DE}. \quad (3)$$

Here, P$_{FE}$ is the polarization induced by ferroelectricity, P$_{FE}$=P$_R$ in full polarization condition. Using SiO$_2$ as DE layer and ε$_{DE}$=3.9, E$_{DE}$ can be estimated as about 30 MV/cm to 90 MV/cm, which is far above the breakdown electric field (E$_{BD}$) of SiO$_2$. Therefore, to satisfy the charge balance condition, there must be charge injection to the FE/DE interface as interfacial charge (Q$_{FEDE}$) through a leakage-assist-switching mechanism during FE polarization switching process in Fe-FET, as shown in Fig. 2(a) and 2(c). The continuity equation becomes,

$$P_{FE} + \epsilon_{FE}E_{FE} = \epsilon_{DE}E_{DE} + Q_{FEDE}. \quad (4)$$

Thus, E$_{DE}$ below E$_{BD}$ can be achieved.

At DE and semiconductor interface, the continuity equation becomes,

$$\epsilon_{DE}E_{DE} = Q_{ch} + Q_{sc} + Q_{it}. \quad (5)$$

where Q$_{ch}$ is the channel carrier density, Q$_{sc}$ is the charge density in space charge region and $Q_{it}$ is the interfacial charge at DE and semiconductor interface. Therefore, to obtain charge balance in Fe-FET in both polarization up and down states, we need both negative charges and positive charges, such as electrons and holes in channel region for charge balance, which may result in partial polarization switching in FE layer.

Therefore, the difference of FE capacitor and Fe-FETs are mainly because of charge balance in Fe-FETs cannot be achieved without introducing non-ideal interfacial charges.

## III. THE IMPACT OF CHARGE BALANCE ON Fe-FET OPERATION

In the above section, the necessity of introducing $Q_{FEDE}$ is discussed. In this section, we will furtherly show by simple derivation that the memory characteristics of Fe-FETs considering FE/DE interfacial charge are fundamentally different from without considering $Q_{FEDE}$. Memory characteristics such as MW, endurance, retention, etc. are determined by DE layer instead of FE layer as commonly understood.

### A. Memory Window

The memory window of a Fe-FET is defined as the threshold voltage ($V_T$) difference between polarization up and polarization down states. The threshold voltage of a Fe-FET without considering ferroelectric polarization ($V_{T0}$) [30] is the same as conventional metal-oxide-semiconductor FET (MOSFET), which can be written as,

$$V_{T0} = \Phi_{ms} + 2\Psi_B + \frac{\sqrt{4\epsilon_S q N_A \Psi_B}}{C_{ox}},$$ (6)

where $\phi_{ms}$ is the work function difference between metal gate and semiconductor, $\Psi_B$ is the Fermi level from intrinsic Fermi level, $\epsilon_S$ is the dielectric constant of semiconductor, $N_A$ is doping concentration, $C_{ox}$ is the oxide capacitance, q is the elementary charge. $C_{ox}$ is the capacitance of FE layer ($C_{FE}$) and capacitance of DE layer ($C_{DE}$) connected in series. $V_{T0}$ may also depends on the substrate voltage, channel thickness, etc. in fully depleted device and other device structures, but is only used as a reference $V_T$ and will not change the essential physics.

If considering ferroelectric polarization and FE/DE interfacial charge, the threshold voltage of a Fe-FET is,

$$V_T = V_{T0} + \frac{P_{FE} - Q_{FEDE}}{C_{FE}}.$$ (7)

Here, $P_{FE}$ is the FE polarization and is positive in polarization up state, which may not reach full polarization to $P_R$. The interfacial charge at FE/DE interface always has the same sign with $P_{FE}$ due to a leakage-assist-switching mechanism. Therefore, assuming a symmetric polarization switching, memory window can be expressed as ($P_{FE}$ here is the positive one),

$$MW = \frac{2(P_{FE} - Q_{FEDE})}{C_{FE}}.$$ (8)

In ideal case without $Q_{FEDE}$, if $P_{FE}=P_R$ and $Q_{FEDE}$ is zero, and assuming the thickness of FE layer ($t_{FE}$) is 10 nm with $\epsilon_{FE}$ of 20 and $P_R$ is 20 $\mu C/cm^2$, MW of a typical FE HfO2 based Fe-FET is expected to be 22 V, which is too large to be realistic. Therefore, $Q_{FEDE}$ has to be introduced in the model of Fe-FET,

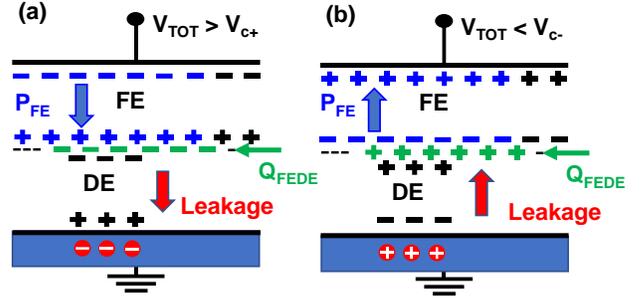

Fig. 3. Leakage-assist-switching in a FE/DE stack in polarization (a) down and (b) up states.

which further confirms the validity of the leakage-assist-switching mechanism.

To properly determine the $Q_{FEDE}$, according to the continuity equation required by charge balance in equation (4), $Q_{FEDE}$ can be approximately written as,

$$Q_{FEDE} = -\epsilon_{DE}E_{eff} + P_{FE}.$$ (9)

Here, ferroelectric polarization is fairly assumed to be much larger than the dielectric components so that the dielectric components can be ignored. $E_{eff}$ is the $E_{DE}$ at sufficiently high gate bias when polarization switching is triggered.

According to the leakage assistant switching mechanism [27] in Fig. 3, $E_{eff}$ can be regarded as an effective electric field that the DE layer becomes leaky so that charge injection to the FE/DE interface becomes possible. This $E_{eff}$ is related with the strength of DE layer and the degradation of DE layer will result in the reduction of $E_{eff}$.

Therefore, considering equations (8) and (9), MW can be written as,

$$MW = 2\frac{\epsilon_{DE}E_{eff}}{C_{FE}}.$$ (10)

In this case, assuming $t_{FE}$ of 10 nm, $\epsilon_{DE}$ of 3.9 and $\epsilon_{FE}$ of 20, $E_{eff}$ of 5 MV/cm, MW can be estimated as 2.0 V, which is very close to experimental reported values.

The above discussions are based on that $P_{FE}$ is larger than $\epsilon_{DE}E_{eff}$ so that a leakage-assist-switching process must exist. However, $P_{FE}$ may not always reach full polarization or high polarization charge density. For example, if gate voltage is not high enough, only partial polarization switching may be obtained, leading to a small memory window. Note that to achieve high $P_{FE}$, both positive and negative charges are required for charge balance according to equation (5). In the semiconductor channel, high positive and negative charge density may not exist together, so that full polarization switching may not be achieved, which may also lead to a small MW.

In this case, if $P_{FE}$ is smaller than $\epsilon_{DE}E_{eff}$, $Q_{FEDE}$ is zero and MW becomes,

$$MW = \frac{2P_{FE}}{C_{FE}}.$$ (11)

Here, according to above new insights, we want to emphasize that there are two key points to achieve a large MW on a Fe-FET. Firstly, the quality of DE layer is critical to enhance the memory window. A low-quality interfacial layer may lead to the diminish of MW because DE layer cannot hold charge. Secondly, to have both high positive and negative charge density, for example high hole and electron density, in semiconductor channel is important for charge balance. Narrow bandgap materials such as Ge [20], [31] and Si may be more suitable for Fe-FETs. AlGaN/GaN heterojunction, where 2-dimensional electron gas formed at its interface, is not suitable for making a Fe-FET because no inversion hole or positive charged layer from heavily doped ions can be formed to support ferroelectric polarization switch in AlGaN/GaN heterojunction. [24], [32]

### B. Endurance

An important indication from above model is to understand the endurance characteristics of FE HfO2 based Fe-FETs. FE HfO2 based M/FE/M capacitors have been demonstrated with high endurance over $10^{12}$ cycles while endurance on the level of only $10^6$ or even below is commonly observed on Fe-FETs with FE HfO2 gate insulator. This phenomenon can be successfully explained by the above model. According to equation (10), MW of Fe-FET degrades during the endurance measurement because $E_{eff}$ becomes smaller. The physics of $E_{eff}$ degradation is simply because $E_{eff}$ by definition is the electric field that DE layer becomes leaky so that charge injection into FE/DE interface becomes possible. Thus, $E_{eff}$ must be close to the breakdown electric field of DE layer. The multiple cycles of operation near breakdown can easily make the DE layer leakier so that $E_{eff}$ becomes smaller. This happens before FE layer degrades. Therefore, the endurance characteristics of a Fe-FET is determined by the DE layer instead of FE layer. This is fundamentally different from a M/FE/M capacitor.

Therefore, to improve the endurance characteristics, there are two possible approaches. First, to develop a FE thin film with FE polarization density less than few $\mu C/cm^2$, so that DE layer does not operate near breakdown. For example, CuInP2S6 has a low $P_R$ of 2-4 $\mu C/cm^2$. [33] The reduction of Hf composition in FE HZO can also reduce the $P_R$ of FE layer. The $P_R$ is also reduced rapidly once FE HZO thickness is smaller than 4 nm. [14] Second, to develop Fe-FETs without DE layer. The control of $D_{it}$ becomes critical for this approach because $D_{it}$ may also screen the FE polarization which leads to the reduction of MW and endurance performance. A back-gate Fe-FET with oxide semiconductor channel [34]–[36] is preferred since no native interfacial oxide layer is needed.

### C. Retention

The retention characteristics are determined by $E_{dep}$ and charge trapping across the DE layer, which both lead to the drift of $V_T$ and lose of MW. More than 10 years retention time has been successfully demonstrated on FE HfO2 based Fe-FETs. The high retention performance is also related with the charge accumulation at FE/DE interface.

According to the definition of $E_{dep}$ (i.e. $E_{dep}t_{FE}+E_{DE}t_{DE}=0$), equations (4) and (9), and a M/FE/DE/M structure for simplicity of discussion, $E_{dep}$ can be written as,

$$E_{dep} = \frac{P_{FE} - Q_{FEDE}}{\epsilon_{FE} + \epsilon_{DE} \frac{t_{FE}}{t_{DE}}}, \qquad (12a)$$

$$E_{dep} \cong \frac{E_{eff}}{\frac{\epsilon_{FE}}{\epsilon_{DE}} + \frac{t_{FE}}{t_{DE}}}. \qquad (12b)$$

As we can see from equation (12a), if assuming $t_{FE}$ of 10 nm, $t_{DE}$ of 1 nm, $\epsilon_{DE}$ of 3.9, $\epsilon_{FE}$ of 20 and $P_R$ of 20 $\mu C/cm^2$, without considering $Q_{FEDE}$, $E_{dep}$ is about 3.8 MV/cm, which is above $E_C$ of FE HfO2 ($\sim$ 1 MV/cm), and it is not stable. If considering $Q_{FEDE}$, according to equation (12b), $E_{dep}$ is 0.3 MV/cm, which is below $E_C$. To increase $\epsilon_{FE}$ for higher dielectric constant and increase the ratio of $t_{FE}$ and $t_{DE}$ can further reduce $E_{dep}$ to improve the retention characteristics.

### D. The Impact of 2D and 3D Electrostatics

The above discussions are based on a one-dimensional (1D) model, but the real Fe-FETs are planar devices with 2D electrostatic potential distribution or 3D devices, such as FinFETs or gate-all-around (GAA) FETs, with 3D electrostatic potential distribution. The impact of 2D or 3D electrostatics is crucial to Fe-FET operation in certain device structures.

For example, in a silicon Fe-FETs with n-type channel (p-doped), to switch down the FE polarization, a high positive gate voltage is applied, so that an electron inversion layer is formed. The high-density electron inversion layer terminates the electric field so that the electrostatic potential distribution is similar to 1D case like a M/FE/DE/M structure as shown in Fig 3(a). To switch up the FE polarization, a high negative gate voltage is applied, so that a hole accumulation layer is formed. The electrostatic potential distribution is also similar to 1D case as shown in Fig 3(b). Therefore, if with both sufficient positive charges and negative charges, Fe-FET operation can be well-explained by the above 1D model.

However, if without the existence of both positive and negative charges, such as a GaN Fe-FET [24], [32], [37] with electrons formed at AlGaN/GaN polar semiconductor interface, Fe-FET switching is not possible due to the lack of holes. There is a large voltage drop across the semiconductor in this case, also resulting in insufficient electric field across the FE layer. This case was verified by numerical simulation in Ref. [24]. Another example is in Fe-FETs with floating body structure [34], [38], [39], there is no well-defined body potential. If without mobile electrons and holes together, the electrostatic potential cannot be approximated as 1D in the off-state, so FE polarization switching near the source/drain region is much easier due to a stronger electric field across the FE layer near source drain. Thus, for Fe-FETs with floating body structure, the performance of short channel devices is expected to be better than long channel devices. Therefore, device scaling can contribute to improve the memory performance of Fe-FETs.

### IV. CONCLUSION

In summary, we have developed a comprehensive Fe-FET model based on the charge balance framework. The impact of charge balance and the leakage-assist-switching mechanism on the memory characteristics of Fe-FETs with M/FE/DE/S gate stack is studied and the discrepancy between the memory performance of FE capacitors and Fe-FETs are successfully understood. It is found that the FE/DE interface and DE layer are critical to determine the memory characteristics of Fe-FETs. The MW, retention and endurance characteristics of FE HfO2 based Fe-FET are determined by DE layer instead of FE layer as usually understood.

To further improve the memory performance of Fe-FETs, according to the new insights in this model, the following two potential approaches are provided. (i) To reduce the remnant polarization of FE layer down to below few $\mu C/cm^2$, so that FE operation does not involve leakage-assist-switching process. For example, FE HZO with a reduced Hf composition or ultrathin FE HZO. (ii) To remove DE layer and use a M/FE/S structure, such as employing a back-gate Fe-FET with an oxide semiconductor channel. But a high-quality FE and semiconductor interface is still needed in this case.

## REFERENCES

[1] H. Ishiwara, "Current status and prospects of FET-type ferroelectric memories," *FED J.*, vol. 11, pp. 27–40, 2000, doi: 10.1109/DRC.1999.806306.

[2] T. S. Böscke, J. Müller, D. Bräuhaus, U. Schröder, and U. Böttger, "Ferroelectricity in hafnium oxide thin films," *Appl. Phys. Lett.*, vol. 99, no. 10, p. 102903, 2011., doi: 10.1063/1.3634052.

[3] J. Müller, T. S. Böscke, U. Schröder, R. Hoffmann, T. Mikolajick, and L. Frey, "Nanosecond polarization switching and long retention in a novel MFIS-FET based on ferroelectric HfO₂," *IEEE Electron Device Lett.*, vol. 33, no. 2, pp. 185–187, 2012, doi: 10.1109/LED.2011.2177435.

[4] J. Müller *et al.*, "Ferroelectricity in Simple Binary ZrO₂ and HfO₂," *Nano Lett.*, vol. 12, no. 8, pp. 4318–4323, 2012, doi: 10.1021/nl302049k.

[5] J. Muller *et al.*, "Ferroelectricity in HfO₂ enables nonvolatile data storage in 28 nm HKMG," in *IEEE Symposium on VLSI Technology*, 2012, pp. 25–26, doi: 10.1109/VLSIT.2012.6242443.

[6] S. Dünkel *et al.*, "A FeFET based super-low-power ultra-fast embedded NVM technology for 22nm FDSOI and beyond," in *IEEE Intl. Electron Devices Meet.*, 2017, pp. 485–488.

[7] K. Ni *et al.*, "Critical Role of Interlayer in Hf₀.₅Zr₀.₅O₂ Ferroelectric FET Nonvolatile Memory Performance," *IEEE Trans. Electron Devices*, vol. 65, no. 6, pp. 2461–2469, 2018, doi: 10.1109/TED.2018.2829122.

[8] K. Florent *et al.*, "First demonstration of vertically stacked ferroelectric Al doped HfO₂ devices for NAND applications," in *IEEE Symposium on VLSI Technology*, 2017, pp. T158–T159, doi: 10.23919/VLSIT.2017.7998162.

[9] K. Florent *et al.*, "Vertical Ferroelectric HfO₂ FET based on 3-D NAND Architecture: Towards Dense Low-Power Memory," in *IEEE Intl. Electron Devices Meet.*, 2018, pp. 43–46, doi: 10.1109/IEDM.2018.8614710.

[10] M.-K. Kim, I. Kim, and J. Lee, "CMOS-compatible ferroelectric NAND flash memory for high-density, low-power, and high-speed three-dimensional memory," *Sci. Adv.*, vol. 7, no. 3, p. eabe1341, 2021, doi: 10.1126/sciadv.abe1341.

[11] H. W. Park, J. Lee, and C. S. Hwang, "Review of ferroelectric field-effect transistors for three-dimensional storage applications," *Nano Sel.*, pp. 1–21, 2021, doi: 10.1002/nano.202000281.

[12] M. H. Park, H. J. Kim, Y. J. Kim, W. Lee, T. Moon, and C. S. Hwang, "Evolution of phases and ferroelectric properties of thin Hf₀.₅Zr₀.₅O₂ films according to the thickness and annealing temperature," *Appl. Phys. Lett.*, vol. 102, p. 242905, 2013, doi: 10.1063/1.4811483.

[13] X. Tian, S. Shibayama, T. Nishimura, T. Yajima, S. Migita, and A. Toriumi, "Evolution of ferroelectric HfO₂ in ultrathin region down to 3 nm," *Appl. Phys. Lett.*, vol. 112, p. 102902, 2018, doi: 10.1063/1.5017094.

[14] X. Lyu, M. Si, X. Sun, M. A. Capano, H. Wang, and P. D. Ye, "Ferroelectric and Anti-Ferroelectric Hafnium Zirconium Oxide: Scaling Limit, Switching Speed and Record High Polarization Density," in *IEEE Symposium on VLSI Technology*, 2019, pp. T44–T45, doi: 10.23919/VLSIT.2019.8776548.

[15] H. Ishiwara, "FeFET and ferroelectric random access memories," *J. Nanosci. Nanotechnol.*, vol. 12, no. 10, pp. 7619–7627, 2012, doi: 10.1093/acprof:oso/9780199584123.003.0012.

[16] N. Gong and T. P. Ma, "Why Is FE-HfO₂ More Suitable Than PZT or SBT for Scaled Nonvolatile 1-T Memory Cell? A Retention Perspective," *IEEE Electron Device Lett.*, vol. 37, no. 9, pp. 1123–1126, 2016, doi: 10.1109/LED.2016.2593627.

[17] E. Yurchuk *et al.*, "Impact of Scaling on the Performance of HfO₂-Based Ferroelectric Field Effect Transistors," *IEEE Trans. Electron Devices*, vol. 61, no. 11, pp. 3699–3706, 2014, doi: 10.1109/TED.2014.2354833.

[18] K. Karda, C. Mouli, and M. A. Alam, "Switching Dynamics and Hot Atom Damage in Landau Switches," *IEEE Electron Device Lett.*, vol. 37, no. 6, pp. 801–804, 2016, doi: 10.1109/LED.2016.2562007.

[19] C. Alessandri, P. Pandey, A. Abusleme, and A. Seabaugh, "Switching Dynamics of Ferroelectric Zr-Doped HfO₂," *IEEE Electron Device Lett.*, vol. 39, no. 11, pp. 1780–1783, 2018, doi: 10.1109/LED.2018.2872124.

[20] W. Chung, M. Si, P. R. Shrestha, J. P. Campbell, K. P. Cheung, and P. D. Ye, "First Direct Experimental Studies of Hf₀.₅Zr₀.₅O₂ Ferroelectric Polarization Switching Down to 100-picosecond in Sub-60mV/dec Germanium Ferroelectric Nanowire FETs," in *IEEE Symposium on VLSI Technology*, 2018, pp. T89–T90, doi: 10.1109/VLSIT.2018.8510652.

[21] X. Lyu, M. Si, P. R. Shrestha, K. P. Cheung, and P. D. Ye, "First Direct Measurement of Sub-Nanosecond Polarization Switching in Ferroelectric Hafnium Zirconium Oxide," in *IEEE Intl. Electron Devices Meet.*, 2019, pp. 342–345, doi: 10.1109/IEDM19573.2019.8993509.

[22] M. Si *et al.*, "Ultrafast measurements of polarization switching dynamics on ferroelectric and anti-ferroelectric hafnium zirconium oxide," *Appl. Phys. Lett.*, vol. 115, no. 7, p. 072107, 2019, doi: 10.1063/1.5098786.

[23] Y. C. Chiu, C. H. Cheng, C. Y. Chang, M. H. Lee, H. H. Hsu, and S. S. Yen, "Low power 1T DRAM/NVM versatile memory featuring steep sub-60-mV/decade operation, fast 20-ns speed, and robust 85°C-extrapolated 10¹⁶ endurance," in *IEEE Symposium on VLSI Technology*, 2015, pp. T184–T185, doi: 10.1109/VLSIT.2015.7223671.

[24] M. Si, Z. Lin, J. Noh, J. Li, W. Chung, and P. D. Ye, "The Impact of Channel Semiconductor on the Memory Characteristics of Ferroelectric Field-Effect Transistors," *IEEE J. Electron Devices Soc.*, vol. 8, pp. 846–849, 2020, doi: 10.1109/JEDS.2020.3012901.

[25] T. P. Ma and J. P. Han, "Why is nonvolatile ferroelectric memory field-effect transistor still elusive?," *IEEE Electron Device Lett.*, vol. 23, no. 7, pp. 386–388, 2002, doi: 10.1109/LED.2002.1015207.

[26] Y. J. Kim *et al.*, "Interfacial charge-induced polarization switching in Al₂O₃/Pb(Zr,Ti)O₃ bi-layer," *J. Appl. Phys.*, vol. 118, no. 22, p. 224105, 2015, doi: 10.1063/1.4937544.

[27] M. Si, X. Lyu, and P. D. Ye, "Ferroelectric Polarization Switching of Hafnium Zirconium Oxide in a Ferroelectric/Dielectric Stack," *ACS Appl. Electron. Mater.*, vol. 1, no. 5, pp. 745–751, 2019, doi: 10.1021/acsaelm.9b00092.

[28] K. Toprasertpong, M. Takenaka, and S. Takagi, "Direct Observation of Interface Charge Behaviors in FeFET by Quasi-Static Split C-V and Hall Techniques : Revealing FeFET Operation," in *IEEE Intl. Electron Devices Meet.*, 2019, vol. 12, pp. 570–573.

[29] R. R. Mehta, B. D. Silverman, and J. T. Jacobs, "Depolarization fields in thin ferroelectric films," *J. Appl. Phys.*, vol. 44, no. 8, pp. 3379–3385, 1973, doi: 10.1063/1.1662770.

[30] S. M. Sze and K. Ng, *Physics of Semiconductor Devices*, 3rd ed. Wiley, 2006.

[31] W. Chung, M. Si, and P. D. Ye, "First Demonstration of Ge Ferroelectric Nanowire FET as Synaptic Device for Online Learning in Neural Network with High Number of Conductance State and Gmax/Gmin," in *IEEE Intl. Electron Devices Meet.*, 2018, pp. 344–347, doi: 10.1109/IEDM.2018.8614516.

[32] C. H. Wu *et al.*, "High V$_{th}$ enhancement mode GaN power devices with high I$_{D, max}$ using hybrid ferroelectric charge trap gate stack," in *IEEE Symposium on VLSI Technology*, 2017, pp. T60–T61, doi: 10.23919/VLSIT.2017.7998201.

[33] M. Si, P.-Y. Y. Liao, G. Qiu, Y. Duan, and P. D. Ye, "Ferroelectric Field-Effect Transistors Based on MoS₂ and CuInP₂S₆ Two-Dimensional van der Waals Heterostructure," *ACS Nano*, vol. 12, no. 7, pp. 6700–6705, 2018, doi: 10.1021/acsnano.8b01810.

[34] M. Si *et al.*, "Indium–Tin-Oxide Transistors with One Nanometer Thick Channel and Ferroelectric Gating," *ACS Nano*, vol. 14, no. 9, pp. 11542–11547, 2020, doi: 10.1021/acsnano.0c03978.

[35] M. Si, A. Charnas, Z. Lin, and P. D. Ye, "Enhancement-Mode Atomic-Layer-Deposited In₂O₃ Transistors With Maximum Drain Current of 2.2 A/mm at Drain Voltage of 0.7 V by Low-Temperature Annealing and Stability in Hydrogen Environment," *IEEE Trans. Electron Devices*, vol. 68, no. 3, pp. 1075–1080, 2021, doi: 10.1109/TED.2021.3053229.

[36] M. Si, Z. Lin, A. Charnas, and P. D. Ye, "Scaled Atomic-Layer-Deposited Indium Oxide Nanometer Transistors With Maximum Drain Current Exceeding 2 A/mm at Drain Voltage of 0.7 V," *IEEE Electron Device Lett.*, vol. 42, no. 2, pp. 184–187, 2021, doi: 10.1109/LED.2020.3043430.

[37] C. Wu *et al.*, "Hf₀.₅Zr₀.₅O₂-Based Ferroelectric Gate HEMTs with Large Threshold Voltage Tuning Range," *IEEE Electron Device Lett.*, vol. 41,

no. 3, pp. 337–340, 2020, doi: 10.1109/LED.2020.2965330.

[38] S. Dutta *et al.*, "Monolithic 3D Integration of High Endurance Multi-Bit Ferroelectric FET for Accelerating Compute-In-Memory," in *IEEE Int. Electron Devices Meet.*, 2020, pp. 801–804, doi: 10.1109/IEDM13553.2020.9371974.

[39] A. A. Sharma *et al.*, "High speed memory operation in channel-last, back-gated ferroelectric transistors," in *IEEE Int. Electron Devices Meet.*, 2020, pp. 391–394, doi: 10.1109/IEDM13553.2020.9371940.